# Routing protocol for wireless quantum multi-hop Mesh backbone network based on partially entangled GHZ state

Pei- Ying Xiong[1], Xu-Tao Yu[1,†], Zai-Chen Zhang[3], Hai-Tao Zhan[1], Jing-Yu Hua[3]

[1] *State Key Lab of Millimeter Waves, Southeast University, Nanjing 210096, China*

[2] *National Mobile Communications Research Laboratory, Southeast University, Nanjing 210096, China*

[3] *College of Information Engineering, Zhejiang University of Technology, Hangzhou 3100231, China*

*Corresponding author. E-mail:* [†]*yuxutao@seu.edu.cn*

Quantum multi-hop teleportation is important in the field of quantum communication. In this study, we propose a quantum multi-hop communication model and a quantum routing protocol with multi-hop teleportation for wireless mesh backbone networks. Based on an analysis of quantum multi-hop protocols, a partially entangled Greenberger--Horne--Zeilinger (GHZ) state is selected as the quantum channel for the proposed protocol. Both quantum and classical wireless channels exist between two neighboring nodes along the route. With the proposed routing protocol, quantum information can be transmitted hop by hop from the source node to the destination node. Based on multi-hop teleportation based on the partially entangled GHZ state, a quantum route established with the minimum number of hops. The difference between our routing protocol and the classical one is that in the former, the processes used to find a quantum route and establish quantum channel entanglement occur simultaneously. The Bell state measurement results of each hop are piggybacked to quantum route finding information. This method reduces the total number of packets and the magnitude of air interface delay. The deduction of the establishment of a quantum channel between source and destination is also presented here. The final success probability of quantum multi-hop teleportation in wireless mesh backbone networks was simulated and analyzed. Our research shows that quantum multi-hop teleportation in wireless mesh backbone networks through a partially entangled GHZ state is feasible.

## 1. Introduction

Quantum communication technology is a popular subject of research in the field of quantum information [1-5]. There are two methods to transmit quantum information between nodes. One involves the direct transfer of quantum states, whereas the other facilitates the indirect transfer of such states through quantum teleportation. Considering that qubits are sensitive to the environment and, hence, that some loss of information can occur, when quantum information needs to be transferred between two distant nodes, quantum teleportation is a more feasible method. Quantum teleportation is thus widely used to transmit quantum information, and quantum entanglement is usually exploited in quantum teleportation. At the beginning of such a transfer, communication resources are usually EPR pairs, named after Einstein, Podolsky, and Rosen. Bennett *et al*. proposed the first protocol for long-distance quantum communication [1], which was experimentally verified in 1997 by Bouwmeester *et al*. [2]. Entanglement swapping helps overcome the distance limitation between two nodes. It introduces intermediate nodes to realize entanglement in point-to point communication. In recent years, partially entangled states have garnered significant interest [6-15]. Sheng *et al*. proposed a scheme for efficient, two-step entanglement concentration based on arbitrary W states [6]. Gour proposed faithful teleportation with partially entangled states [7]. Dai *et al*. presented a scheme where a partially entangled Greenberger--Horne--Zeilinger (GHZ) state and a partially entangled W state are used as the quantum channel to teleport an unknown two-particle state from a sender to either of two receivers [8].

In the future, long-distance transmission of quantum information through networks will emerge as an important field of research. With the development of quantum communication, research on multi-hop teleportation is required [16-20]. In order to transmit quantum information between nodes that do not share direct entanglement, intermediate nodes are usually introduced. Quantum channels can be built through entanglement shared by adjacent nodes. Shi *et al*. proposed a quantum wireless multi-hop network where quantum information is transmitted hop by hop through teleportation, and W states are used as entangled resource [16]. Cai *et al*. proposed a quantum bridging method to teleport qubits using partially entangled states in hop-by-hop transmission [17]. Wang *et al*. [19] proposed and verified a method to help teleport a quantum state from source to destination. In this method, intermediate nodes can implement teleportation in parallel. In our previous work, we proposed a quantum multi-hop teleportation protocol based on a partially entangled GHZ state [20].

While wireless technology has provided flexible methods for classical communication networks, it is rarely used in quantum communication networks even though there is exciting potential for application. Few studies have addressed complex, wireless quantum communication networks [19-23]. Wang *et al*. proposed a quantum routing protocol that can reduce end-to-end communication delay by using simultaneous measurements [19]. Cheng *et al*. proposed a quantum routing mechanism using quantum teleportation in a hierarchical network architecture [20]. Yu *et al*. proposed a routing protocol for a wireless ad-hoc quantum communication network (WAQCN) [21].

In this paper, we describe our proposed model for a wireless quantum multi-hop communication network with a mesh backbone structure. In this network, we introduce a special node, called the edge route node, to connect the client node with the backbone. By studying existing quantum multi-hop protocols, we selected a partially entangled GHZ state as the entangled resource: we have studied this in [18] in the context of quantum channels. The proposed routing protocol is different from the classical routing protocol in that the processes of quantum route finding and quantum channel establishment occurs simultaneously in ours. In the route finding process, the Bell state measurement results of each hop are piggybacked to the quantum route finding packet. This routing protocol reduces the total number of packets and the magnitude of the air interface delay.

The rest of this article is organized as follows: In Section 2, we describe our design of a mesh structure in wireless quantum networks. Section 3 is devoted to an analysis of prevalent research on quantum multi-hop teleportation. We also revise our past insights on quantum multi-hop teleportation based on a partially entangled GHZ state. In Section 4, a quantum communication routing protocol based on the partially entangled GHZ state is proposed for multi-hop teleportation in wireless mesh networks. In Section 5, quantum channel establishment in the proposed routing protocol is deduced. An example of routing transmission is presented in Section 6. The corresponding success probability of quantum teleportation in this network is analyzed and simulation results are provided in Section 7. Finally, we draw our conclusions in Section 8.

## 2. A mesh structure in wireless quantum multi-hop communication

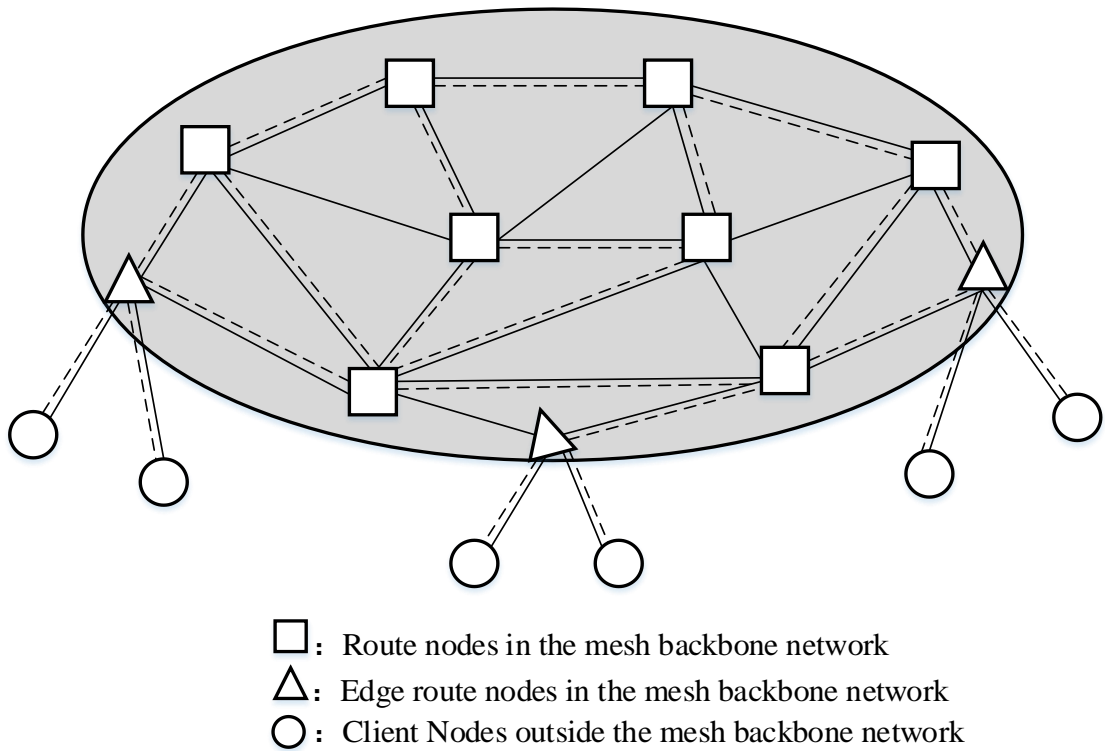

□：Route nodes in the mesh backbone network
△：Edge route nodes in the mesh backbone network
○：Client Nodes outside the mesh backbone network

**Fig.1** A model of the wireless quantum communication network

The wireless mesh network (WMN) is a self-organizing and self-healing network that has been employed for classical wireless communication. It is an easily scalable and low-cost network that uses a multi-hop routing mechanism and a hierarchy configuration. The WMN consists of mesh routers and mesh clients. Mesh routers are stationary or have minimal mobility, and serve as the backbone of the WMN in providing wireless access to mesh clients. Mesh clients are removable.

Based on the traditional WMN, a mesh structure is introduced to wireless quantum communication, and a model of this wireless quantum mesh network is shown in Fig. 1. The figure shows that there are three kinds of nodes and two kinds of channels. The gray oval represents the scope of the mesh backbone network, similar to mesh routers in the traditional WMN. The backbone network is composed of route nodes and edge route nodes. The square nodes in the gray oval are route nodes that do not directly connect to client nodes. Triangular nodes, called edge route nodes, act as bridging points to implement quantum and wireless access between nodes in the mesh backbone network and client nodes outside the network. The circular nodes represent client nodes that have a direct quantum and wireless channel with a neighboring edge route node. Every edge route node has a client list to record addresses of client nodes connected to it.

There are two kinds of communication channels in this distributed wireless quantum network: classical wireless and quantum wireless channels. The solid lines in Fig. 1 represent classical communication channels and the dotted lines represent quantum communication channels. Classical information such as measurement results and Bell pair types are transmitted through classical wireless channels. Only when classical and quantum routes exist simultaneously can quantum information be transmitted between nodes. Nodes along the quantum route share entanglement one by one, and quantum information is transferred through the quantum route. Nodes along the classical route can transfer classical information. For two nodes, if there is a direct classical or a quantum route between them, they are called neighbor nodes in quantum or wireless, respectively. For client nodes connected to the same edge route node, quantum information can be teleported directly between them or through their edge route node. For nodes not connected with the same edge route node, a quantum route through the backbone network can be selected to realize long-distance quantum communication by introducing intermediate route nodes. With this method, quantum information can be transferred hop by hop from source nodes to destination nodes.

### 3. Related works on quantum multi-hop channels

Since there is no possibility that a source node shares entangled particles with every other node, a multi-hop teleportation protocol is necessary for quantum communication. With the help of multi-hop teleportation, the source node can transfer quantum information to intermediate nodes hop by hop through quantum communication channels in proper order. Finally, quantum information would be transferred to the destination node. Quantum channels between neighbor nodes can be built with the aid of an entangled resource. In light of the instability of maximally quantum states, partially entangled quantum states are a better choice for entangled resource. Researchers in the literature [14-18] have selected partially entangled EPR pairs, W states, and partially entangled GHZ states as entangled resource to establish quantum multi-hop channels. Following the analysis of the success probability of each quantum multi-hop teleportation protocol in related work, we find that success probability increases as the number of hops decreases. Therefore, we select routes with the minimum number of hops in the wireless quantum multi-hop communication network as far as possible. In this article, we select partially entangled GHZ states as entangled resource to establish quantum communication channels.

In our previous work [18], a quantum multi-hop teleportation scheme based on the partially entangled GHZ state was proposed. The overall success rate after the $i$-th teleportation via auxiliary particles was deduced as follows:

$$P_{total}^{(i)} = 2 \sum_{j=0}^{(i-1)/2} \binom{i}{j} \frac{n^{2(i-j)}}{(1+n^2)^i}, \qquad i \text{ is odd}$$

$$P_{total}^{(i)} = \binom{i}{i/2} \frac{n^i}{(1+n^2)^i} + \sum_{j=1}^{i/2} \binom{i}{\frac{i}{2}-j} \frac{2n^{i+2j}}{(1+n^2)^i}, \qquad i \text{ is even} \tag{1}$$

where $i$ represents the number of quantum teleportations and $n$ represents the degree of entanglement of the partially entangled GHZ state. Looking at Eq. (1), the probability of successful teleportation based on $i$ and $n$. Setting $n$ as a constant value, the relationship between success probability and the instances of teleportation $i$ can be obtained. When $n$ =0.5, 0.7, 0.9, $i$ varies from 1 to 150; the probability of successful teleportation is presented in Fig. 2. From Fig. 2, we find that the possibility of successful teleportation is deeply influenced by the number of instances of teleportation $i$. The higher the value of $i$, the lower the success probability. In our study, in order to obtain a higher likelihood of successful teleportation in the quantum mesh backbone network, the route with the least nodes will be selected.

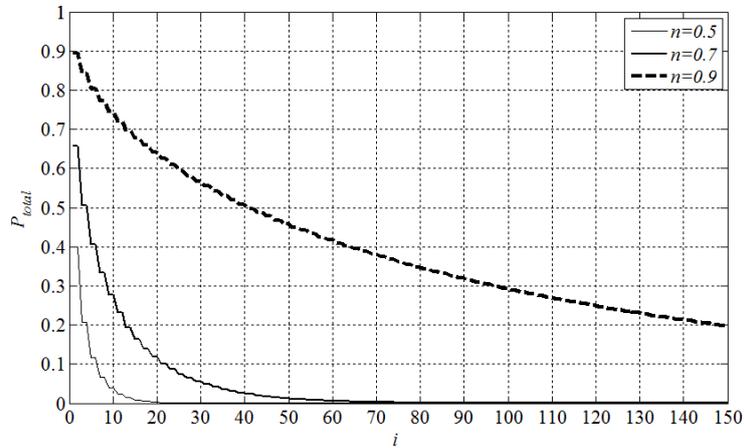

**Fig. 2** The possibility of successful teleportation

## 4. Routing protocol

The simplified quantum mesh backbone network is shown in Fig. 3. The solid lines represent classical communication channels and the dotted ones depict quantum communication channels. Considering that there is no possibility that the source node shares an entangled resource with every other node, intermediate nodes are needed to transfer quantum information between source and destination nodes. As shown in Fig. 3, source node A does not share an entangled resource directly with destination node

H, and entanglement swapping can be used to establish quantum channels between them. Measurement results are transferred through classical channels. The quantum and classical routes can be different; only when the two simultaneously coexist can the quantum state be transferred between two client nodes. To this end, a routing protocol to select the quantum route in quantum communication networks is proposed. The routing protocol consists of two parts: quantum route request (QRR) and quantum route finding (QRF).

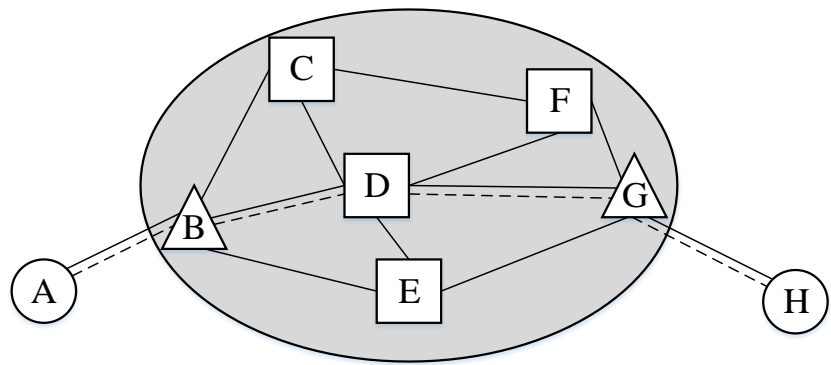

**Fig.5** Quantum circuit for N nodes.

The simplified quantum mesh backbone network is given in Fig. 5, solid lines represent quantum communication channels and dotted lines represent classical communication channels. The entanglement in this quantum mesh backbone network is really precious and consumed in every-hop quantum teleportation process. Therefore, it is no possibility that source node A shares entangled resource with every other node. As shown in Fig. 5, source node A does not share entangled resource with destination node H and entanglement swapping can be utilized to establish quantum channels between them. Measurement results are transferred in classical channels. Only when classical and quantum channels are existed simultaneously, quantum state can be transferred between two nodes. In this part, a routing protocol in quantum communication network is proposed. Quantum paths and classical paths can be different. We call two nodes that there are quantum and classical channels between them as neighbor nodes. Neighbor nodes share a pair of partially entangled GHZ particles which are able to guarantee quantum communication. Entanglement swapping helps to transfer quantum information hop by hop from source node A to destination H. The routing protocol consists of these parts: quantum route setup (QRS) and quantum route find (QRF).

### 4.1 Quantum route request

In Fig. 3, when client node A wants to send information to destination node H, it checks to determine if there is an available route to node H. If there is none, node A sends a quantum route request (QRR) packet to the neighboring edge route node B through the wireless channel. The QRR packet contains the source node, the destination node, route request message ID, the address of the current node, and the value of the routing cost. The initial value of the routing cost is set to 0. When A sends a new routing request message, the ID needs to be incremented by one. The address of the current node is the address of A.

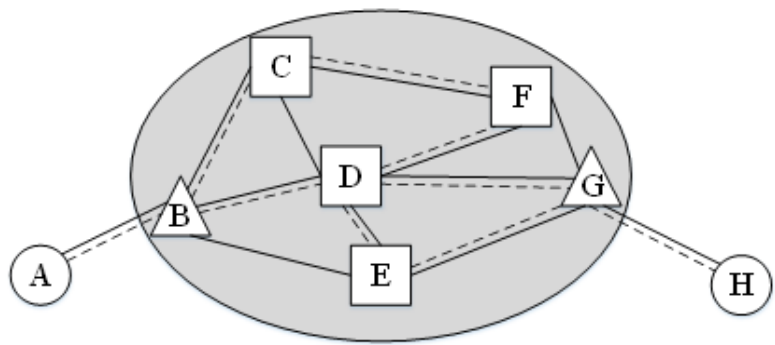

**Fig.3** The simplified quantum mesh backbone network.

Once the neighboring edge route node B receives the QRR, it checks whether it has received the same QRR by checking its ID and source node address, following which the route cost in QRR adds 1. The route table is updated; it contains the upward address, the downward address, the ID, the value of the routing cost, the source node, and the destination node. The downward address, the ID, the value of the routing cost, and the source and destination nodes are identical to the corresponding information in the received QRR. At the same time, we update the address of the previous node in the QRR with that of the current node and broadcast the QRR in the quantum communication backbone network. Node B then broadcasts the QRR to the backbone network on a wireless channel.

The intermediate node in the backbone network receiving the QRR checks whether it has received the same QRR by checking the ID and source node address. If these are identical, the QRR is discarded; otherwise, the route cost in QRR adds 1, and the node broadcasts the updated QRR. The edge route node in the backbone network receiving the QRR checks whether the address of destination node is in its client list. If not, the QRR is discarded. Otherwise, the process of quantum route finding is initiated.

**4.2 Quantum route finding**

Once the edge route node, which neighbors the destination node, receives the QRR, the route finding process begins. This process includes route selection and route reply. The hop-by-hop node address is recorded in the received QRR. According to these address records, edge route node G selects an appropriate route according to the routing cost, and adds the selected route to the route table. The route satisfying the following conditions is selected: the quantum and wireless routes exist simultaneously, and route cost is minimal. If there is more than one route satisfying these conditions, the edge route node G chooses the one whose quantum route request is first received by G. Once the route is selected, the neighboring edge route node of the destination sends a reply message to the source node by the reverse route. When the reply of quantum route finding begins, quantum channel establishment begins as well. Edge route node G performs Bell state measurement on two of its particles, which belong to two pairs of partially entangled GHZ states. Node G then performs the Hadamard gate operation on its remaining particles, measures the state of these particles, and adds the measurement result and the Bell state measurement results to the reply message of the QRF. According to the reverse path of the selected route, G sends reply message QRF to edge route B in unicasting mode, where the route response message QRF includes the node addresses of the selected route and the measurement results. When nodes along the selected route, including the edge route node, receive the QRF, they update the route table according to the route information in the received QRF. Then performs Bell state measurement on two of its particles belonging to two pairs of partially entangled GHZ states, performs the Hadamard gate operation on its remaining particles, and measures the state of these. The measurement results are added to the QRF and sent to the next node along the selected route. Finally, the QRF is sent to source node A though edge route node B.

When source node A receives the QRF, a direct quantum channel between the source and destination nodes is established, which means that the quantum state can be directly teleported from source to destination. Therefore, the source node performs Bell state measurement on two particles. Then, through the selected route, the source node transmits a result packet to a neighbor of the destination node through the selected route. The result packet includes measurement results at the source node and those in the received QRF. Edge route node G then transfers the result packet to destination node H. Once the destination node receives the result packet, it deals with the results and chooses proper operations to recover the transmitted quantum state. This quantum communication process is hence successfully completed. Later in this article, the specific deduction of quantum channel establishment is provided.

**4.3   Comparison with other routing protocols**

Wang *et al.* used simultaneous Bell measurements in all intermediate nodes [17]. This scheme reduces delays in teleportation to a greater extent than the sequential entanglement swapping scheme. In this scheme, measurements at each node along the selected route can be simultaneously carried out and measurement results can be sent to the destination node independently, which reduces delays in hop-by-hop teleportation.

In our protocol, in contrast to the method proposed in [17], the quantum channel between source and destination is established by a sequential entanglement swapping scheme that leads to more delays. On the contrary, in our protocol, the swapping process is combined with quantum route finding, and the measurement results are piggybacked to QRF packets. Therefore, the results need not be transmitted through specified packets, and hence the total number of packets in the network decreases, which reduces network delay in turn.

**5.   The process of establishment of the quantum channel**

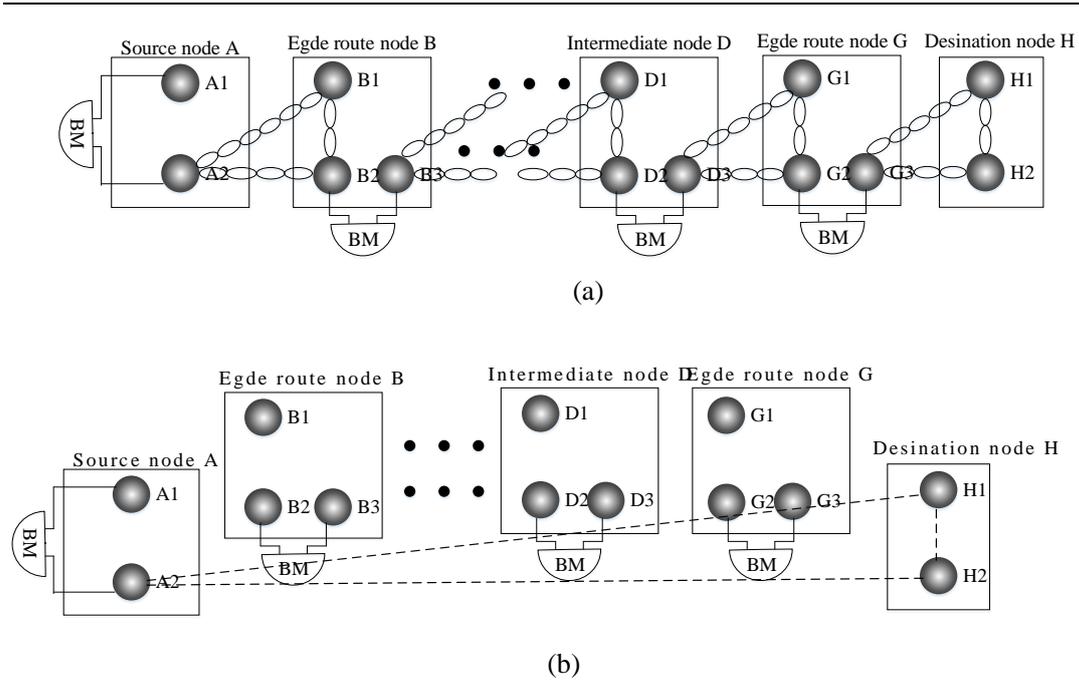

**Fig. 4** Quantum channel establishment process:

(a) Before quantum route finding process, (b) After quantum route finding process.

Quantum channel establishment prior to quantum route finding in our routing protocol is shown in Fig. 4(a). The source and destination nodes are A and H, respectively. Source node A contains an unknown state $|\psi\rangle_{A1} = \alpha|0\rangle_{A1} + \beta|1\rangle_{A1}$. Assume that between A and H, there are $N$-1 nodes. Neighboring nodes share a partially entangled GHZ state $|GHZ\rangle = \frac{1}{\sqrt{(1+n^2)}}(|000\rangle + n|111\rangle)$ as quantum channel. The edge route node G first performs Bell state measurement on particles G2 and G3, and a Hadamard gate operation on particle G1. The states of all particles can be written as

$$
\begin{aligned}
|\psi\rangle &= |\psi\rangle_{A1} \otimes_{i=1}^{N} |GHZ_{n_i}\rangle = |\psi\rangle_{A1} \otimes \frac{1}{(1+n^2)} \\
&\times \{|\phi^+\rangle_{G2G3}[|0\rangle_{G1}(|000\rangle_{D3H1H2} + n^2|111\rangle_{D3H1H2}) + |1\rangle_{G1}(|000\rangle_{D3H1H2} - n^2|111\rangle_{D3H1H2})] \\
&+ |\phi^-\rangle_{G2G3}[|0\rangle_{G1}(|000\rangle_{D3H1H2} - n^2|111\rangle_{D3H1H2}) + |1\rangle_{G1}(|000\rangle_{D3H1H2} + n^2|111\rangle_{D3H1H2})] \\
&+ |\Psi^+\rangle_{G2G3}[|0\rangle_{G1}n(|001\rangle_{D3H1H2} + |110\rangle_{D3H1H2}) - |1\rangle_{G1}n(|001\rangle_{D3H1H2} - |110\rangle_{D3H1H2})] \\
&+ |\Psi^-\rangle_{G2G3}[|0\rangle_{G1}n(|001\rangle_{D3H1H2} - |110\rangle_{D3H1H2}) + |1\rangle_{G1}n(|001\rangle_{D3H1H2} + |110\rangle_{D3H1H2})]\} \\
&\otimes_{i=2}^{N}|GHZ_{n_i}\rangle
\end{aligned}
\quad (2)
$$

In Eq. (2), we set $0 < n_i < 1$ and $n_i = n$, $i = 1,...,N$

The node then applies the proper Pauli operator on particle H2. The quantum state can then be rewritten as

$$|\psi\rangle = |\psi\rangle_{A1} \otimes_{i=1}^{N} |GHZ_{n_i}\rangle = |\psi\rangle_{A1} \otimes \frac{1}{(1+n^2)}$$

$$\times \{|\phi^+\rangle_{G2G3}[|0\rangle_{G1}(|000\rangle_{D3H1H2} + n^2|111\rangle_{D3H1H2}) + |1\rangle_{G1}\sigma_z(|000\rangle_{D3H1H2} + n^2|111\rangle_{D3H1H2})]$$

$$+ |\phi^-\rangle_{G2G3}[|0\rangle_{G1}\sigma_z(|000\rangle_{D3H1H2} + n^2|111\rangle_{D3H1H2}) + |1\rangle_{G1}(|000\rangle_{D3H1H2} + n^2|111\rangle_{D3H1H2})]$$

$$+ |\Psi^+\rangle_{G2G3}[|0\rangle_{G1}\sigma_x n(|000\rangle_{D3H1H2} + |111\rangle_{D3H1H2}) - |1\rangle_{G1}\sigma_z\sigma_x n(|000\rangle_{D3H1H2} + |111\rangle_{D3H1H2})]$$

$$+ |\Psi^-\rangle_{G2G3}[|0\rangle_{G1}\sigma_z\sigma_x n(|000\rangle_{D3H1H2} + |111\rangle_{D3H1H2}) + |1\rangle_{G1}\sigma_x n(|000\rangle_{D3H1H2} + |111\rangle_{D3H1H2})]\}$$

$$\otimes_{i=2}^{N} |GHZ_{n_i}\rangle \quad (3)$$

When $N=1$, node A is the neighbor of node H. Particles A2, H1, and H2 are partially entangled GHZ states. We perform Bell measurement on particles A1 and A2, and apply the Pauli operator to particle H2. If the Bell measurement result is $|\phi^{\pm}\rangle$, the system state is given by

$$|\psi\rangle = |\psi\rangle_{A1} \otimes |\psi\rangle_{A2H1H2}$$
$$= (\alpha|0\rangle_{A1} + \beta|1\rangle_{A1})(|000\rangle_{A2H1H2} + n|111\rangle_{A2H1H2})$$
$$= \frac{1}{2} \times \{|\phi^+\rangle_{A1A2}[|0\rangle_{H1}(\alpha|0\rangle_{H2} + n\beta|1\rangle_{H2}) + |1\rangle_{H1}\sigma_z(\alpha|0\rangle_{H2} + n\beta|1\rangle_{H2})] \quad (4)$$
$$+ |\phi^-\rangle_{A1A2}[|0\rangle_{H1}\sigma_z(\alpha|0\rangle_{H2} + n\beta|1\rangle_{H2}) + |1\rangle_{H1}(\alpha|0\rangle_{H2} + n\beta|1\rangle_{H2})]$$
$$+ |\Psi^+\rangle_{A1A2}[|0\rangle_{H1}\sigma_x(n\alpha|0\rangle_{H2} + \beta|1\rangle_{H2}) - |1\rangle_{H1}\sigma_z\sigma_x(n\alpha|0\rangle_{H2} + \beta|1\rangle_{H2})]$$
$$+ |\Psi^-\rangle_{A1A2}[|0\rangle_{H1}\sigma_z\sigma_x(n\alpha|0\rangle_{H2} + \beta|1\rangle_{H2}) - |1\rangle_{H1}\sigma_x(n\alpha|0\rangle_{H2} + \beta|1\rangle_{H2})]\}$$

Then, using the method of analysis that introduces an auxiliary and performs proper unitary matrix, $|\psi\rangle_{H2} = \alpha|0\rangle_{H2} + \beta|1\rangle_{H2}$ can be obtained at destination node H. If the result of the Bell measurement is $|\Psi^{\pm}\rangle$, we can obtain $|\psi\rangle_{H2} = \alpha|0\rangle_{H2} + \beta|1\rangle_{H2}$ in a similar manner.

Assume that quantum finding information is transferred to node $X^j$, which is along the selected route, and $j$ Bell measurements have been carried out, where $j = 1,...,N$. Destination node H is equivalent to $X^0$ and node G is equivalent to $X^1$. Prior to performing the Bell measurement at edge route node G, the state of edge route nodes G and H is $|000\rangle + n|111\rangle$. Before performing Bell measurement at node $X^j$, we assume that the state of the upward node and H is $m_{(j-1)}|000\rangle + k_{(j-1)}|111\rangle$; when the measurement result is $|\phi^{\pm}\rangle$, the state of node $X^j$ and H is $m_{(j-1)}|000\rangle + nk_{(j-1)}|111\rangle$. When the measurement result is $|\Psi^{\pm}\rangle$, the state of node $X^j$ and H is $nm_{(j-1)}|000\rangle + k_{(j-1)}|111\rangle$. Assume the state of node $X^j$ and H is $m_j|000\rangle + k_j|111\rangle$; perform Bell measurement at source node A, which is equivalent to $X^N$, and quantum channel establishment after quantum route finding process is presented in Fig. 4(b). The dotted lines represent quantum entanglement channels. The states of A and H are given by

$$|\psi\rangle = |\psi\rangle_{A1} \otimes |\psi\rangle_{A2H1H2}$$
$$= (\alpha|0\rangle_{A1} + \beta|1\rangle_{A1})(m_N|000\rangle_{A2H1H2} + k_N|111\rangle_{A2H1H2})$$
$$= \frac{1}{2} \times \{|\phi^+\rangle_{A1A2}[|0\rangle_{H1}(m_N\alpha|0\rangle_{H2} + k_N\beta|1\rangle_{H2}) + |1\rangle_{H1}\sigma_z(m_N\alpha|0\rangle_{H2} + k_N\beta|1\rangle_{H2})] \quad (5)$$
$$+ |\phi^-\rangle_{A1A2}[|0\rangle_{H1}\sigma_z(m_N\alpha|0\rangle_{H2} + k_N\beta|1\rangle_{H2}) + |1\rangle_{H1}(m_N\alpha|0\rangle_{H2} + k_N\beta|1\rangle_{H2})]$$
$$+ |\Psi^+\rangle_{A1A2}[|0\rangle_{H1}\sigma_x(k_N\alpha|0\rangle_{H2} + m_N\beta|1\rangle_{H2}) - |1\rangle_{H1}\sigma_z\sigma_x(k_N\alpha|0\rangle_{H2} + m_N\beta|1\rangle_{H2})]$$
$$+ |\Psi^-\rangle_{A1A2}[|0\rangle_{H1}\sigma_z\sigma_x(k_N\alpha|0\rangle_{H2} + m_N\beta|1\rangle_{H2}) - |1\rangle_{H1}\sigma_x(k_N\alpha|0\rangle_{H2} + m_N\beta|1\rangle_{H2})]\}$$

By introducing an auxiliary and performing a proper unitary matrix, we can obtain $|\psi\rangle_{H2} = \alpha|0\rangle_{H2} + \beta|1\rangle_{H2}$ at destination node H. When $m_N < k_N$, the state of particle H2 is $\alpha|0\rangle_{H2} + n^{m_N/k_N}\beta|1\rangle_{H2}$. The corrected process is as follows:

$$U\{(\alpha|0\rangle_{H2} + n^{m_N/k_N}\beta|1\rangle_{H2}) \otimes |0\rangle_{aux}\}$$
$$= n^{m_N/k_N}(\alpha|0\rangle_{H2} + \beta|1\rangle_{H2})|0\rangle_{aux} \quad (6)$$
$$+ \sqrt{1-(n^{m_N/k_N})^2}\alpha|1\rangle_{H2}|1\rangle_{aux}$$

where the unitary matrix is given by

$$U = \begin{pmatrix} n^{m_N/k_N} & \sqrt{1-(n^{m_N/k_N})^2} & 0 & 0 \\ 0 & 0 & 0 & 1 \\ 0 & 0 & 1 & 0 \\ \sqrt{1-(n^{m_N/k_N})^2} & -n^{m_N/k_N} & 0 & 0 \end{pmatrix} \quad (7)$$

When $m_N > k_N$, the state of particle H2 is $n^{k_N/m_N}\alpha|0\rangle_{H2} + \beta|1\rangle_{H2}$, The corrected process is as follows:

$$V\{(\alpha|0\rangle_{H2} + n^{k_N/m_N}\beta|1\rangle_{H2}) \otimes |0\rangle_{aux}\}$$
$$= n^{k_N/m_N}(\alpha|0\rangle_{H2} + \beta|1\rangle_{H2})|0\rangle_{aux} \quad (8)$$
$$+ \sqrt{1-(n^{k_N/m_N})^2}\alpha|1\rangle_{H2}|1\rangle_{aux}$$

where the unitary matrix is given by

$$V = \begin{pmatrix} 1 & 0 & 0 & 0 \\ 0 & 0 & 0 & 1 \\ 0 & \sqrt{1-(n^{k_N/m_N})^2} & n & 0 \\ 0 & -n & \sqrt{1-(n^{k_N/m_N})^2} & 0 \end{pmatrix} \quad (9)$$

The success probability of quantum establishment is given by

$$P_{total}^{(i)} = 2 \sum_{j=0}^{(i-1)/2} \binom{i}{j} \frac{n^{2(i-j)}}{(1+n^2)^i}, \quad i \text{ is odd}$$

$$P_{total}^{(i)} = \binom{i}{i/2} \frac{n^i}{(1+n^2)^i} + \sum_{j=1}^{i/2} \binom{i}{\frac{i}{2}-j} \frac{2n^{i+2j}}{(1+n^2)^i}, \quad i \text{ is even}$$

(10)

where $i$ is the number of Bell state measurements. This success probability is identical to that given by Eq. (1), which was deduced in our previous work [18].

In Table 1, we present the relevant Pauli operator on particle H2 according to the measurement results of every hop along the selected route.

**Table 1** Relevant Pauli operator on particle H2 according to Bell measurement results and the state of particle performed a Hadamard gate

| Bell measurement result | The state of particle which is performed a Hadamard gate | Pauli operator $\sigma$ on particle H2 |
|---|---|---|
| $\|\phi^+\rangle$ | $\|0\rangle_3$ | Do nothing |
| | $\|1\rangle_3$ | $\sigma_z$ |
| $\|\phi^-\rangle$ | $\|0\rangle_3$ | $\sigma_z$ |
| | $\|1\rangle_3$ | Do nothing |
| $\|\Psi^+\rangle$ | $\|0\rangle_3$ | $\sigma_x$ |
| | $\|1\rangle_3$ | $\sigma_z \sigma_x$ |
| $\|\Psi^-\rangle$ | $\|0\rangle_3$ | $\sigma_z \sigma_x$ |
| | $\|1\rangle_3$ | $\sigma_x$ |

## 5  An example of routing transmission

As shown in Fig. 5(a), source node A seeks to teleport an unknown particle state to destination node H. Node A is not adjacent to node H. The solid paths represent classical communication channels and the dotted paths quantum communication channels. Source node A sends a QRR packet to edge route node B, which includes the addresses of A and H. When B receives the QRR packet, it checks whether destination node H is in its client list. If it is not, the address of B is added to the QRR and the route cost is incremented by one. Then, B broadcast the updated QRR packet in the backbone network. Route nodes C, D, and E receive the QRR packet. They check the packet ID and the source address, and ensure that the packet is the first received. These route nodes then update QRR packets and broadcast them. Route nodes B, C, D, E, F, and G receive the updated QRR packet. By checking the packet ID and source address, nodes B, C, D, and E find that this packet has been received, and discard the received QRR packet. Assume node F receives a QRR packet from node C first. F updates this packet and broadcasts it. It then receives a QRR packet from D, and discards it after checking its ID and source address. Assume edge route node G receives a QRR packet from node D first; G updates this packet. It then receives a QRR packet from E and F, and discards these packets after checking the ID and source address. Thus, edge route node G chooses route A-B-D-G-H. In the direction of route transmission, we define D as the upstream node of G and H as the downstream node of G. Edge node G needs to send a QRF packet to source node A in the reverse direction. The reverse route from G to A is G-D-B-A. For the sake of consistency, we still define D as the upstream of G. The reverse route is established as shown in Fig. 5(b). Edge node G sends a QRF packet that includes the route from source node A to destination node H and the measurement results. When G's upstream node D receives a QRF packet, it continues to send it to its upstream node. When source node A receives a QRF packet, it adds the new measurement results to a result packet, which includes the measurement results of every hop. These are transmitted to destination node H. Operations are carried out at Node H to recover the transmitted quantum information.

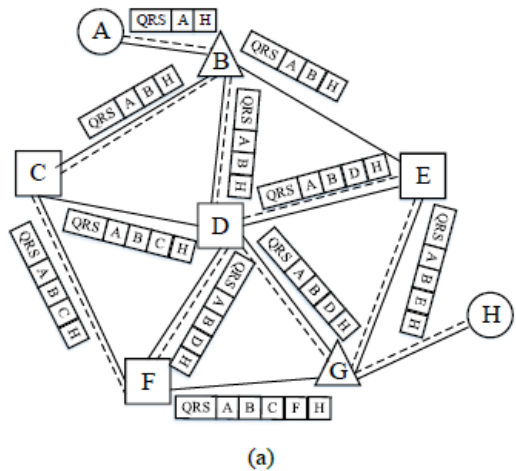

(a)

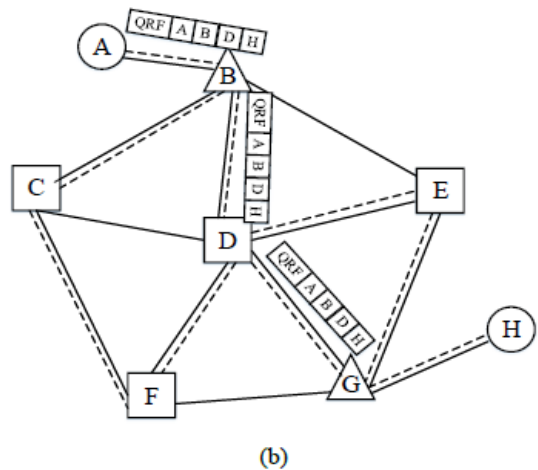

(b)

**Fig. 5** Example of routing transmission: (a) QRR packets of quantum route; (b) QRF packets of quantum route.

## 6    Success probability of quantum route establishment

Based on the success probability of quantum route establishment, the probability $P_{suc}$ of successful quantum teleportation in the mesh backbone network is analyzed. Assume that the area occupied by the network is 1000 m × 1000 m. Nodes in the network are distributed at random. When the number of nodes is a constant value, transmission range $R$ and the probability $p$ of successfully established communication channels between two neighboring nodes are relevant to $P_{suc}$ in the wireless quantum network.

In this work, a mesh structure is introduced in a wireless quantum network, and classical and quantum channels are built simultaneously. We assumed that the success probability of establishing these two kinds of channels is the same. When the distance between two neighboring nodes is no more than the transmission range $R$, information can be transmitted successfully through wireless channels. At the same time, quantum channels exist. Otherwise, quantum teleportation from the source to destination node fails. Assume that edge route nodes are the nearest nodes in the backbone network connected to client nodes. If there are no edge route nodes, quantum teleportation from source to destination, which are not directly connected with each other, fails. Quantum route establishment is based on a minimum number of hops. The success probability of quantum channel establishment via partially entangled GHZ particles is given by Eq. (10). By changing transmission range $R$, the probability $p$ of successfully establishing communication channels, and the number of nodes in the mesh backbone network, the corresponding success probability $P_{suc}$ of quantum teleportation is shown in the following figures. In order to avoid an accident, we let the simulation program run 100 times, and simulation results obtained were the average results.

In simulation Fig. 6, we selected fixed $R$, $R$ =200 m, and changed the probability $p$, $p$ =0.3, 0.5, 0.8. It was obvious that success probability $P_{suc}$ increased with probability $p$, and the number of nodes in the wireless quantum network. In Fig. 7, we set $R$ =300 m and $p$ =0.5, and found that the success probability $P_{suc}$ increased with transmission range $R$. This was because with the increase in network connectivity, the number of elective routes increased, and the corresponding minimal number of hops decreased. In these figures, $n$ represents the degree of entanglement. When $n$ =1, neighboring nodes shared maximally entangled GHZ states. The $P_{suc}$ of the maximally entangled GHZ state was higher than the $P_{suc}$ of a partially entangled GHZ state.

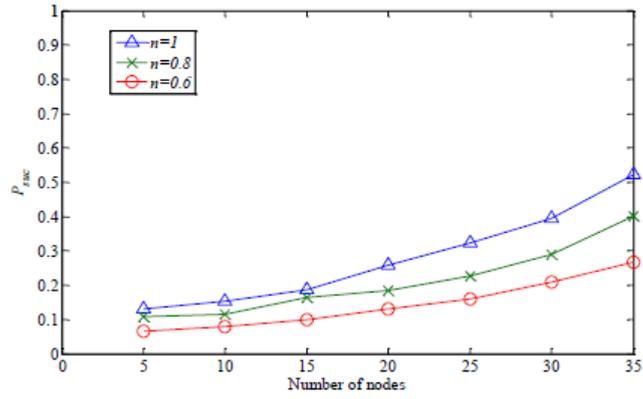

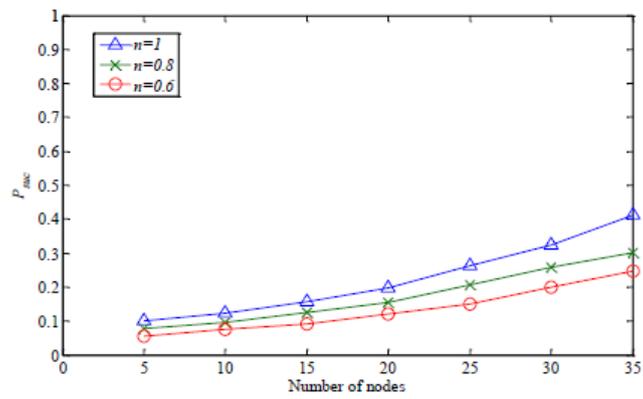

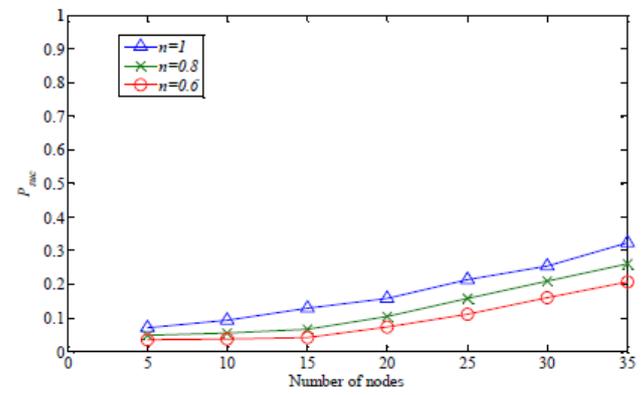

**Fig.6** Success probability of quantum teleportation in the wireless quantum network:
(a) $R=200m$, $p=0.8$, (b) $R=200m$, $p=0.5$, (c) $R=200m$, $p=0.3$.

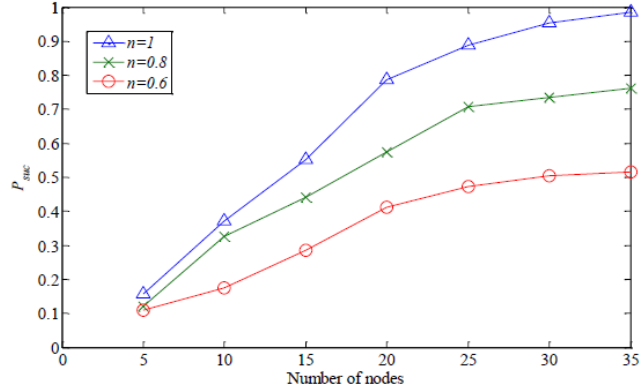

**Fig.7** Success probability of quantum teleportation in the wireless quantum network ($R$=300m, $p$=0.5)

## 7  Conclusions

In this work, we proposed a quantum communication network model where a mesh backbone network structure was introduced. In this structure, edge route nodes acted as bridging nodes to help source nodes establish communication with nodes in the mesh backbone network.

We proposed a quantum routing protocol with multi-hop teleportation for the wireless mesh network with partially entangled GHZ states. Based on our previous protocol for multi-hop partially entangled GHZ state teleportation, the number of hops was selected as route metric. In this routing protocol, only when classical and quantum channels simultaneously co-exist can quantum information be transferred between nodes. The edge route node, which is the neighbor of the destination node, selects the route with the minimum number of hops as the most appropriate route. A direct quantum channel between source and destination was established when the quantum route finding message was transferred to the source node. Quantum channel establishment via partially entangled GHZ states was performed. Based on all measurement results, we conclude that the quantum state can be recovered in destination nodes using a corresponding matrix operation. By comparing the results of our method with the simultaneous entanglement swapping scheme, we found that although quantum measurements are performed hop by hop in our protocol, the measurement results are piggybacked to QRR packets, which reduces the total number of packets transmitted in the wireless channel and reduces network delay. Moreover, the success probability of quantum route establishment was simulated. We saw that the success probability increased with the number of nodes in a certain range; when the number of nodes exceeded some value, the probability tended to stabilize.

**Acknowledgments** This project was supported by the National Natural Science Foundation of China (Grant No. 61571105 and No.61601120), the Prospective Future Network Project of the Jiangsu Province, China (Grant No. BY2013095-1-18), and the Independent Project of State Key Laboratory of Millimeter Waves (Grant No. Z201504).

**References**

1. C. H. Bennett, G. Brassard, C. Crepeau, R. Jozsa, A. Peres, and W. K. Wootters, Teleporting an unknown quantum state via dual classical and Einstein-Podolsky-Rosen channels, *Phys. Rev. Lett.* 70(13), 1895 (1993)

2. D. Bouwmeester, J. W. Pan, K. Mattle, M. Eibl, H. Weinfurter, and A. Zeilinger, Experimental quantum teleportation, *Nature* 390(6660), 575 (1997)

3. D. Bouwmeester, K. Mattle, J. W. Pan, H. Weinfurter, A. Zeilinger, and M. Żukowski, Experimental quantum teleportation of arbitrary quantum states, *Appl. Phys. B* 67(6), 749 (1998)

4. M. Żukowski, A. Zeilinger, M. A. Horne, and A. K. Ekert, "Event-ready-detectors" Bell experiment via entanglement swapping, *Phys. Rev. Lett.* 71(26), 4287 (1993)

5. J. W. Pan, D. Bouwmeester, H. Weinfurter, and A. Zeilinger, Experimental entanglement swapping: Entangling photons that never interacted, *Phys. Rev. Lett.* 80(18), 3891 (1998)

6. Y. B. Sheng, L. Zhou, and S. M. Zhao, Efficient twostep entanglement concentration for arbitrary W states, *Phys. Rev. A* 85(4), 042302 (2012)

7. G. Gour, Faithful teleportation with partially entangled states, *Phys. Rev. A* 70(4), 042301 (2004)

8. H. Y. Dai, P. X. Chen, and C. Z. Li, Probabilistic teleportation of an arbitrary two-particle state by a partially entangled three-particle GHZ state and W state, *Opt. Commun.* 231(1–6), 281 (2004)

9. Y. H. Wang and H. S. Song, Preparation of partially entangled W state and deterministic multi-controlled teleportation, *Opt. Commun.* 281(3), 489 (2008)

10. Z. Kurucz, M. Koniorczyk, and J. Janszky, Teleportation with partially entangled states, *Fortschr. Phys.* 49(10–11), 1019 (2001)

11. D. P. Tian, Y. J. Tao, and M. Qin, Teleportation of anarbitrary two-qudit state based on the non-maximally four-qudit cluster state, *Sci. China Ser. G-Phys. Mech. Astron.* 51(10), 1523 (2008)

12. N. B. An, Probabilistic teleportation of an M-quNit state by a single non-maximally entangled quNit-pair, *Phys. Lett. A* 372(21), 3778 (2008)

13. G. Rigolin, Unity fidelity multiple teleportation using partially entangled states, *J. Phys. At. Mol. Opt. Phys.* 42(23), 235504 (2009)

14. M. Jiang, H. Li, Z. K. Zhang, and J. Zeng, Faithful teleportation of multi-particle states involving multi spatially remote agents via probabilistic channels, *Physica A* 390(4), 760 (2011)

15. M. Jiang, H. Li, Z. K. Zhang, and J. Zeng, Faithful teleportation via multi-particle quantum states in a network with many agents, *Quantum Inform. Process.* 11(1), 23 (2012)

16. L. H. Shi, X. T. Yu, X. F. Cai, Y. X. Gong, and Z. C. Zhang, Quantum information transmission in the quantum wireless multihop network based on Werner state, *Chin. Phys. B* 24(5), 050308 (2015)

17. X. F. Cai, X. T. Yu, L. H. Shi, and Z. C. Zhang, Partially entangled states bridge in quantum teleportation, *Front. Phys.* 9(5), 646 (2014)

18. X. T. Yu, J. Xu, and Z. C. Zhang, Distributed wireless quantum communication

networks, *Chin. Phys. B* 22(9), 090311 (2013)

19. K. Wang, X. T. Yu, S. L. Lu, and Y. X. Gong, Quantum wireless multi-hop communication based on arbitrary Bell pairs and teleportation, *Phys. Rev. A* 89(2), 022329 (2014)

20. P. Y. Xiong, X. T. Yu, H. T. Zhan, and Z. C. Zhang, Multiple teleportation via partially entangled GHZ state, *Front. Phys.* 11(4), 110303 (2016)

21. K. Wang, Y. X. Gong, X. T. Yu, and S. L. Lu, Addendum to "Quantum wireless multihop communication based on arbitrary Bell pairs and teleportation", *Phys. Rev. A* 90(4), 044302 (2014)

22. S. T. Cheng, C. Y. Wang, and M. H. Tao, Quantum communication for wireless wide-area networks, *IEEE J. Sel. Areas Comm.* 23(7), 1424 (2005)

23. X. T. Yu, J. Xu, and Z. C. Zhang, Routing protocol for wireless ad hoc quantum communication network based on quantum teleportation, *Acta Phisica Sinica* 61(22), 220303 (2012)

24. X. T. Yu, Z. C. Zhang, and J. Xu, Distributed wireless quantum communication networks with partially entangled pairs, *Chin. Phys. B* 23(1), 010303 (2014)

25. I. F. Akyildiz, X. Wang, and W. Wang, Wireless mesh networks: A survey, *Comput. Netw. ISDN Syst.* 47(4), 445 (2005)